\documentclass[preprint,preprintnumbers, prd, floatfix, superscriptaddress,nofootinbib] {revtex4-1}
\usepackage{epsfig}
\usepackage{subfigure}
\usepackage{amsmath}
\usepackage{dcolumn}
\usepackage{bm}
\usepackage[usenames ,dvipsnames]{xcolor}
\usepackage{slashed}
\usepackage{graphicx,color}
\usepackage[bookmarksnumbered, pdfstartview=FitH,colorlinks,urlcolor=blue, citecolor=blue,linkcolor=blue] {hyperref}
\begin{document}
\title{Amplitude analysis for charmed meson decays at BESIII}

\author{Han Zhang}
\affiliation{School of Physics and Microelectronics,
Zhengzhou University, Zhengzhou, Henan 450001, China}

\author{Zhenxuan Li}
\affiliation{Nankai University, Tianjin 300071, China}

\author{Chunyi Guan}
\email{Corresponding author: guancy@ihep.ac.cn}    
\affiliation{Institute of High Energy Physics, Beijing 100049, China}

\author{Zehui Lu}
\affiliation{Institute of High Energy Physics, Beijing 100049, China}

\author{Hui Li}
\affiliation{Nankai University, Tianjin 300071, China}

\author{Liping Yang}
\affiliation{Institute of High Energy Physics, Beijing 100049, China}

\author{Yechun Yu}
\affiliation{School of Nuclear Science and Technology, University of
  Chinese Academy of Sciences, Beijing 101408, China}

\author{Nan Zhang}
\affiliation{Jilin University, Changchun 130012, People’s Republic of China}

\author{Minggang Zhao}
\affiliation{Nankai University, Tianjin 300071, China}

\author{Yu Lu}
\affiliation{Central South University, Changsha 410083, China}

\author{Bai-Cian Ke}
\affiliation{School of Physics and Microelectronics,
Zhengzhou University, Zhengzhou, Henan 450001, China}

\author{Liaoyuan Dong}
\affiliation{Institute of High Energy Physics, Beijing 100049, China}

\date{\today}

\begin{abstract}
Amplitude analysis bridges the gap between experimental measurements of multibody charmed-meson decays and theoretical predictions of intermediate two-body processes. This work presents a comprehensive overview of the amplitude-analysis methodology employed by the BESIII Collaboration, emphasizing practical implementation. We detail the construction of the probability density function and likelihood function for unbinned maximum-likelihood fits. This encompasses Monte Carlo integration techniques for normalization, the incorporation of detection efficiency and resolution effects, and multidimensional background modeling utilizing XGBoost classifiers. Furthermore, we describe the amplitude formalism for both hadronic and semileptonic decays, incorporating standard resonance-propagator parametrizations. Key analytical aspects, including the evaluation of fit fractions, the generation of kinematic projections, and the estimation of statistical uncertainties, are also discussed.
\end{abstract}

\maketitle
\section{introduction}
The necessity of amplitude analysis stems from the inherent disparity between experimental capabilities and theoretical formulations in particle physics. Experimentally, typical detectors directly observe only stable or sufficiently long-lived particles, such as $e^\pm$, $\mu^\pm$, $\pi^\pm$, $K^\pm$, $p$, and $\gamma$. Other common final states, such as $K_S^0$ and $\pi^0/\eta$, are reconstructed via their decays \textcolor{red}{ $K_S^0 \to \pi\pi$} and $\pi^0/\eta \to \gamma\gamma$, respectively. These constitute the experimentally accessible final-state particles. Conversely, short-lived resonances, such as $K^*$, $\phi$, and \textcolor{red}{$a_0/f_0$~\cite{PDG, Achasov:2017edm}}, decay promptly and evade direct detection. Theoretical frameworks, however, are largely agnostic to the stability of the decay products. \textcolor{red}{Nevertheless}, due to the nonperturbative nature of the strong interaction, rigorous theoretical predictions for multibody decays remain highly challenging and are often constrained to two-body or quasi-two-body intermediate processes.

Decay modes yielding three or more final-state particles inherently exhibit quantum interference among intermediate resonant states. For instance, the final state of $D^0 \to K^- \pi^+ \pi^0$ receives interfering contributions from intermediate processes such as $D^0 \to \bar{K}^{*0} \pi^0 \to K^- \pi^+ \pi^0$ and $D^0 \to K^{*-} \pi^+ \to K^- \pi^+ \pi^0$\textcolor{red}{~\cite{PDG}}. While experiments measure the kinematic phase space (four-momenta) of the final-state $K^-$, $\pi^+$, and $\pi^0$, theoretical calculations are primarily tractable for the quasi-two-body transitions $D^0 \to \bar{K}^{*0} \pi^0$ and $D^0 \to K^{*-} \pi^+$. \textcolor{red}{Therefore, extracting the underlying intermediate dynamics from final-state kinematics, while rigorously accounting for quantum interference, is crucial for testing theoretical models.} Amplitude analysis provides the essential mathematical framework to bridge this experimental-theoretical divide.

\textcolor{red}{The BESIII experiment, located at the Beijing Electron Positron Collider II, operates as a dedicated $\tau$-charm factory. Over the past decade, BESIII has accumulated an unprecedentedly large data sample of $e^+e^-$ collisions at a center-of-mass energy of $\sqrt{s}=3.773$~GeV, reaching an integrated luminosity of $20.3~\text{fb}^{-1}$~\cite{BESIII:2024lbn}. In addition, the experiment has collected about $7.33~\text{fb}^{-1}$ of data in the center-of-mass energy range between 4.128 and 4.226~GeV. The threshold production of $D\bar{D}$ and $D_s^*\bar{D_s}$ pairs provides a uniquely clean experimental environment. The $D_{(s)}$ and $\bar{D_{(s)}}$ mesons are produced nearly at rest, with only the $D_{(s)}\bar{D_{(s)}}$ pair and no additional hadrons, leading to low background and high detection efficiency. These features make BESIII an ideal laboratory for studying $D$ meson decays.}

The remainder of this paper is organized as follows. Section II details the construction of the probability density function and the likelihood function employed in amplitude analyses, encompassing Monte Carlo~(MC) integration techniques, the treatment of detection efficiency and experimental resolution, and background modeling. Section III outlines the amplitude formalism governing the decay dynamics of $D$ mesons as implemented in BESIII measurements. Finally, Section IV provides a summary and explores advanced applications of these methodologies.

\section{Probability Density Function and Likelihood}
In the amplitude analysis of a multibody decay---typically involving three or more final-state particles---the relative magnitudes and phases of intermediate decay processes are extracted via fits to data samples. Other properties, such as resonance masses and widths, may also be treated as free parameters if required. This section details the construction of the likelihood and probability density functions~(PDFs) based on amplitude models for unbinned maximum-likelihood fits.

The signal PDF, representing the probability density of a specific kinematic configuration $p$, is defined as
\begin{equation}
f_S(p) = \frac{\epsilon(p)|\mathcal{M}(p)|^2R(p)}{\int \epsilon(p)|\mathcal{M}(p)|^2R(p)\mathrm{d}p},
\label{pwa:pdf}
\end{equation}
where $\epsilon(p)$ represents the detection efficiency, $R(p)$ is the phase-space~(PHSP) factor, and $p$ denotes the set of kinematic variables characterizing a decay event. The total amplitude $\mathcal{M}(p)$ is the coherent sum of the amplitudes corresponding to intermediate processes, given by
\begin{equation}\label{eq:total_amplitude}
  \mathcal{M}(p) = \sum_n c_n\mathcal{A}_n(p),
\end{equation}
where $c_n = \rho_n e^{i\phi_n}$ and $\mathcal{A}_n$ are the complex coefficient and the dynamic amplitude for the $n^{\mathrm{th}}$ intermediate process, respectively. The magnitude $\rho_n$ and phase $\phi_n$ are free parameters in the fit. The formalism of the individual amplitudes will be detailed in Sec.~\ref{sec:formalism}. 

While the amplitude $\mathcal{M}(p)$ isolates the pure decay dynamics of a $D$ meson independent of detector effects, the experimental data are inevitably subject to a nonuniform detection efficiency. To construct a PDF that accurately models the measured event distribution, the efficiency $\epsilon(p)$ must be incorporated as a multiplicative factor. The set of kinematic variables $p$ typically comprises the four-momenta of the final-state particles. \textcolor{red}{For the decay of a spin-0 mother particle (e.g., a $D$ meson) into $N\geq 3$ particles, the number of degrees of freedom is $3N-7$. This is derived from the $3N$ momentum components, subtracting $4$ constraints from energy-momentum conservation and $3$ Euler angles that define the overall spatial orientation of the final-state system. Owing to the isotropic nature of the decay in the rest frame, these three angles can be ignored.} Further details are available in Chapter 49 of the Particle Data Group~(PDG) review~\cite{PDG}. The PHSP factor $R(p)$ encodes the kinematic phase-space density; its functional form depends on the specific choice of coordinates. It remains constant over the allowed PHSP boundary when parameterized directly in terms of the four-momenta, but may vary in other coordinate representations. Its analytical form derives from the Jacobian determinant associated with the coordinate transformation. Furthermore, the integral in the denominator ensures that the signal PDF is strictly normalized to unity over the entire PHSP, fulfilling the fundamental mathematical requirement of a PDF.

The likelihood for a given dataset is constructed as the product of the PDF evaluated at each measured event:
\begin{equation}
\mathcal{L} = \prod_{k=1}^{N_{\mathrm{data}}} f_S(p_k)\,,
\end{equation}
where $k$ runs over all events in the data sample, and $N_{\mathrm{data}}$ is the total number of events. Consequently, the log-likelihood function, which is maximized during the fitting procedure, is given by
\begin{align}
  \ln\mathcal{L} &= \sum_{k=1}^{N_{\mathrm{data}}} \ln f_S(p_k)\nonumber\\
  &=\sum_{k=1}^{N_{\mathrm{data}}} \ln \frac{|\mathcal{M}(p_k)|^2}{\int \epsilon(p)|\mathcal{M}(p)|^2R(p)\mathrm{d}p} \nonumber \\
  &\quad + \sum_{k=1}^{N_{\mathrm{data}}} \ln \left[\epsilon(p_k) R(p_k)\right]\,.
\end{align}
Because the term $\sum \ln[\epsilon(p_k) R(p_k)]$ is independent of the fit parameters, it acts as a constant offset and can be omitted during the maximization process. Parameter estimation is entirely driven by the first term. Furthermore, the normalization integral in the denominator can be efficiently approximated via MC integration, a technique detailed in Sec.~\ref{sec:MC}. This reveals an elegant feature of the amplitude-analysis formalism: the parameter extraction can be performed without requiring a priori analytical knowledge of the explicit efficiency and PHSP functions.

In the presence of non-negligible background contributions, the likelihood is extended by incorporating a normalized background shape $\mathcal{B}(p)$:
\begin{align}
  \ln\mathcal{L} &= \sum_{k=1}^{N_{\mathrm{data}}} \ln \left[w_{\mathrm{sig}}f_S(p_k)+(1-w_{\mathrm{sig}})\frac{\mathcal{B}(p_k)}{\int \mathcal{B}(p)\mathrm{d}p}\right]\,,
\end{align}
where $w_{\mathrm{sig}}$ denotes the signal purity of the data sample. By defining an efficiency- and PHSP-corrected background PDF as $\mathcal{B}_{\epsilon}(p) = \mathcal{B}(p)/[\epsilon(p) R(p)]$, the term $\epsilon(p_k) R(p_k)$ can again be factored out. The modified log-likelihood then becomes
\begin{align}\label{eq:likelihood_SB}
  \ln\mathcal{L} &= \sum_{k=1}^{N_{\mathrm{data}}} \ln \left[ \frac{w_{\mathrm{sig}}|\mathcal{M}(p_k)|^2}{\int \epsilon(p)|\mathcal{M}(p)|^2R(p)\mathrm{d}p} \right. \nonumber \\ 
  &\quad \left. + \frac{(1-w_{\mathrm{sig}})\mathcal{B}_{\epsilon}(p_k)}{\int \epsilon(p)\mathcal{B}_{\epsilon}(p)R(p)\mathrm{d}p} \right] + \sum_{k=1}^{N_{\mathrm{data}}} \ln \left[\epsilon(p_k) R(p_k)\right]\,.
\end{align}
As before, the additive $\ln[\epsilon R]$ term is dropped during the fit. The corrected background shape $\mathcal{B}_{\epsilon}(p)$ is typically obtained through multidimensional reweighting techniques (discussed in Sec.~\ref{sec:bkg}), utilizing the background distribution $\mathcal{B}(p)$ modeled from inclusive MC samples or data-driven sideband estimations.

An alternative strategy to handle backgrounds is to subtract their contribution directly from the log-likelihood function using simulated or control events:
\begin{align}
  \ln\mathcal{L} &= \frac{-N_{\mathrm{data}}+w N_{\mathrm{bkg}}}{N_{\mathrm{data}}+w^2 N_{\mathrm{bkg}}}\left[\sum_{k=1}^{N_{\mathrm{data}}} \ln f_{S}(p_k) - \sum_{l=1}^{N_{\mathrm{bkg}}} \ln f_{S}(p_l)\right]\,,
\end{align}
where $l$ iterates over events in a dedicated background sample, $N_{\mathrm{bkg}}$ is the total number of such background events, and the statistical scaling weight $w = (1-w_{\mathrm{sig}})N_{\mathrm{data}}/N_{\mathrm{bkg}}$ ensures proper normalization according to the signal purity. The prefactor ensures the correct estimation of statistical uncertainties. However, this background-subtraction approach can lead to numerical instabilities in low-purity regimes and may potentially introduce biases; consequently, the direct background modeling approach is generally preferred.

\subsection{Monte Carlo integration, detection, and resolution}\label{sec:MC}
The normalization integral in the denominator of Eq.~(\ref{eq:likelihood_SB}) can be evaluated via MC integration using a PHSP MC sample~\cite{geant4}. A PHSP MC sample is generated with a uniform decay amplitude while strictly adhering to the kinematic constraints of the decay. Consequently, the kinematic distribution of events in this sample inherently encodes the PHSP density. Summing over this sample automatically accounts for the PHSP factor $R(p)$. The normalization integral is thus approximated as
\begin{equation}
\begin{aligned}
  \int \epsilon(p)|\mathcal{M}(p)|^2R(p)\mathrm{d}p &\approx \frac{V}{N_{\mathrm{gen}}}\sum_{k=1}^{N_{\mathrm{gen}}}\epsilon^{\prime}(p_k)|\mathcal{M}(p_k)|^2,
\end{aligned}
\end{equation}
where $k$ is the event index, $N_{\mathrm{gen}}$ is the total number of generated MC events, and $V=\int R(p)\mathrm{d}p$ represents the total volume of the allowed PHSP. In the analytical integral, $\epsilon(p)$ acts as a continuous efficiency probability function. In the MC evaluation, $\epsilon(p)$ is replaced by a binary indicator $\epsilon^{\prime} \in \{0, 1\}$. Each generated event contributes $|\mathcal{M}(p)|^2$ to the sum with probability $\epsilon(p)$, or is otherwise discarded. 

This binary efficiency is naturally implemented by passing the generated MC sample through full detector simulation and reconstruction algorithms. Each event is either retained or rejected based on the reconstruction criteria. This effectively transforms the sum over generated events into a sum over purely reconstructed events:
\begin{equation}
  \begin{aligned}
  \int \epsilon(p)|\mathcal{M}(p)|^2R(p)\mathrm{d}p &\approx \frac{V}{N_{\mathrm{gen}}}\sum_{k=1}^{N_{\mathrm{rec}}}|\mathcal{M}(p_k^{\mathrm{rec}})|^2,
\end{aligned}
\end{equation}
where $N_{\mathrm{rec}}$ is the number of reconstructed MC events, and $p_k^{\mathrm{rec}}$ denotes the reconstructed kinematics of the $k^{\mathrm{th}}$ event.

While PHSP MC integration is theoretically unbiased, it is computationally inefficient. A PHSP sample is uniformly populated across the allowed kinematic phase space, whereas experimental data typically exhibit pronounced resonant structures; certain kinematic regions contain high data densities while others remain sparse. Integration via a uniform PHSP sample allocates equivalent computational effort regardless of a region's actual contribution. Consequently, for a fixed $N_{\mathrm{gen}}$, computational resources are wasted in sparsely populated regions while failing to achieve adequate sampling precision in densely populated resonant peak regions.

To optimize computational efficiency, an importance-sampling technique is employed utilizing a ``signal MC'' sample. This sample is generated such that its density roughly follows the physical data distribution. In practice, it is obtained by performing a preliminary fit to the data using a PHSP MC sample for normalization, and subsequently generating events distributed according to the fitted amplitude model. Using a signal MC sample, the normalization integral evaluates as 
\begin{equation}\label{eq:signaMCint}
\begin{aligned}
  \int \epsilon(p)|\mathcal{M}(p)|^2R(p)\mathrm{d}p &\approx \frac{1}{N_{\mathrm{gen}}}\sum_{k=1}^{N_{\mathrm{rec}}}\frac{|\mathcal{M}(p_k^{\mathrm{rec}})|^2}{|\mathcal{M}^{\mathrm{gen}}(p_k^{\mathrm{rec}})|^2}\,,
\end{aligned}
\end{equation}
where $\mathcal{M}^{\mathrm{gen}}$ is the dynamic amplitude used to generate the signal MC sample.

Beyond computational efficiency, the signal MC approach provides an elegant mechanism for incorporating detector resolution effects. While the amplitude squared $|\mathcal{M}|^2$ describes the pure physical dynamics, experimental data are inevitably smeared by finite detector resolution. Consequently, intrinsic narrow resonant peaks are broadened in the measured spectra. For resonances with natural widths smaller than a few tens of MeV (such as the $\phi$ meson), modeling this resolution effect is crucial.

Resolution effects are naturally accounted for in Eq.~(\ref{eq:signaMCint}) by utilizing the reconstructed kinematic variables, $p^{\mathrm{rec}}$, rather than the true generator-level variables. Because the reconstructed signal MC sample used for the summation has already undergone the full simulation of detector smearing, evaluating the ratio $|\mathcal{M}(p^{\mathrm{rec}})|^2/|\mathcal{M}^{\mathrm{gen}}(p^{\mathrm{rec}})|^2$ automatically folds the effective resolution smearing into the likelihood. Consider a narrow peak that undergoes detector broadening, taking the ratio of the smeared distribution to the true distribution yields a bimodal or ``m''-shaped weighting curve. Applying this empirical weight $w_m$ during the MC integration effectively smears the theoretical amplitude squared $|\mathcal{M}|^2$.



The normalization integral for the background term in Eq.~(\ref{eq:likelihood_SB}) must be evaluated consistently with the signal methodology (further detailed in Sec.~\ref{sec:bkg}). It is imperative that the signal and background terms in Eq.~(\ref{eq:likelihood_SB}) are integrated over the exact same MC sample footprint; otherwise, relative differences in the normalization constants ($N_{\mathrm{gen}}$) cannot be factorized out and will distort the log-likelihood minimization.

Alternatively, detector resolution can be explicitly modeled by numerically convolving $|\mathcal{M}|^2$ with a resolution function, typically a Gaussian. However, due to the severe computational complexity of multidimensional convolutions, this approach is practically restricted to one-dimensional projections where ultra-narrow resonances dominate. For dimensions lacking such fine structures, resolution effects are generally negligible compared to the intrinsic resonance widths.

\subsection{Background}\label{sec:bkg}
The treatments of the signal shape $|\mathcal{M}|^2$ and the corrected background shape $\mathcal{B}_\epsilon$ within the likelihood function differ fundamentally. Unlike $|\mathcal{M}|^2$, which is parameterized utilizing theoretical amplitude models (see Sec.~\ref{sec:formalism}), $\mathcal{B}_\epsilon$ lacks a first-principles analytical description. In practice, the experimental background comprises a complex mixture of misidentified particles, combinatorial artifacts, and partially reconstructed decays from numerous channels, making a purely analytical derivation impossible. In this section, we outline the extraction of $\mathcal{B}_\epsilon$ utilizing a multidimensional reweighting technique based on an XGBoost classifier~\cite{Liu:2019huh, XGboost}.

As a robust binary classifier, XGBoost discriminates between two classes (A and B), assigning an event $x$ a probability $P_{\mathrm{A}}(x)$ of belonging to class A, with the complementary probability being $P_{\mathrm{B}}(x) = 1 - P_{\mathrm{A}}(x)$. To evaluate $\mathcal{B}_\epsilon$, the classifier is trained using a fully reconstructed PHSP MC sample and a dedicated background MC sample. Because the original PHSP MC events are generated with a uniform decay amplitude, their kinematic distribution post-reconstruction inherently maps the efficiency and phase-space profile, $\epsilon(p)R(p)$. Consequently, according to the principles of density ratio estimation, the output odds ratio $P_{\mathrm{BKG}}(p)/P_{\mathrm{PHSP}}(p)$ serves as an empirical proxy for the ratio of the background density to the $\epsilon(p) R(p)$ distribution, which is mathematically equivalent to $\mathcal{B}_\epsilon(p)$. One can therefore determine the normalization integral for the background term in Eq.~(\ref{eq:likelihood_SB}) by summing the odds ratio over a generated signal MC sample:
\begin{equation}
  \int \epsilon(p)\mathcal{B}_\epsilon(p) R(p)\mathrm{d}p \propto \frac{1}{N_{\mathrm{gen}}}\sum_{k=1}^{N_{\mathrm{rec}}}\frac{\left[\frac{P_{\mathrm{BKG}}(p_k^{\mathrm{rec}})}{P_{\mathrm{PHSP}}(p_k^{\mathrm{rec}})}\right]}{|\mathcal{M}^{\mathrm{gen}}(p_k^{\mathrm{rec}})|^2}\,.
\end{equation}
Accordingly, the normalized background probability evaluated for the $i^{\mathrm{th}}$ measured data event, $p_i$, becomes
\begin{equation}
  \frac{\mathcal{B}_{\epsilon}(p_i)}{\int \epsilon(p)\mathcal{B}_{\epsilon}(p)R(p)\mathrm{d}p} = \frac{\left[\frac{P_{\mathrm{BKG}}(p_i)}{P_{\mathrm{PHSP}}(p_i)}\right]}{\frac{1}{N_{\mathrm{gen}}}\sum_{k=1}^{N_{\mathrm{rec}}}\frac{\left[\frac{P_{\mathrm{BKG}}(p_k^{\mathrm{rec}})}{P_{\mathrm{PHSP}}(p_k^{\mathrm{rec}})}\right]}{|\mathcal{M}^{\mathrm{gen}}(p_k^{\mathrm{rec}})|^2}}\,.
\end{equation}

\begin{figure}[htp!]
  \centering
  \includegraphics[width=3.0in]{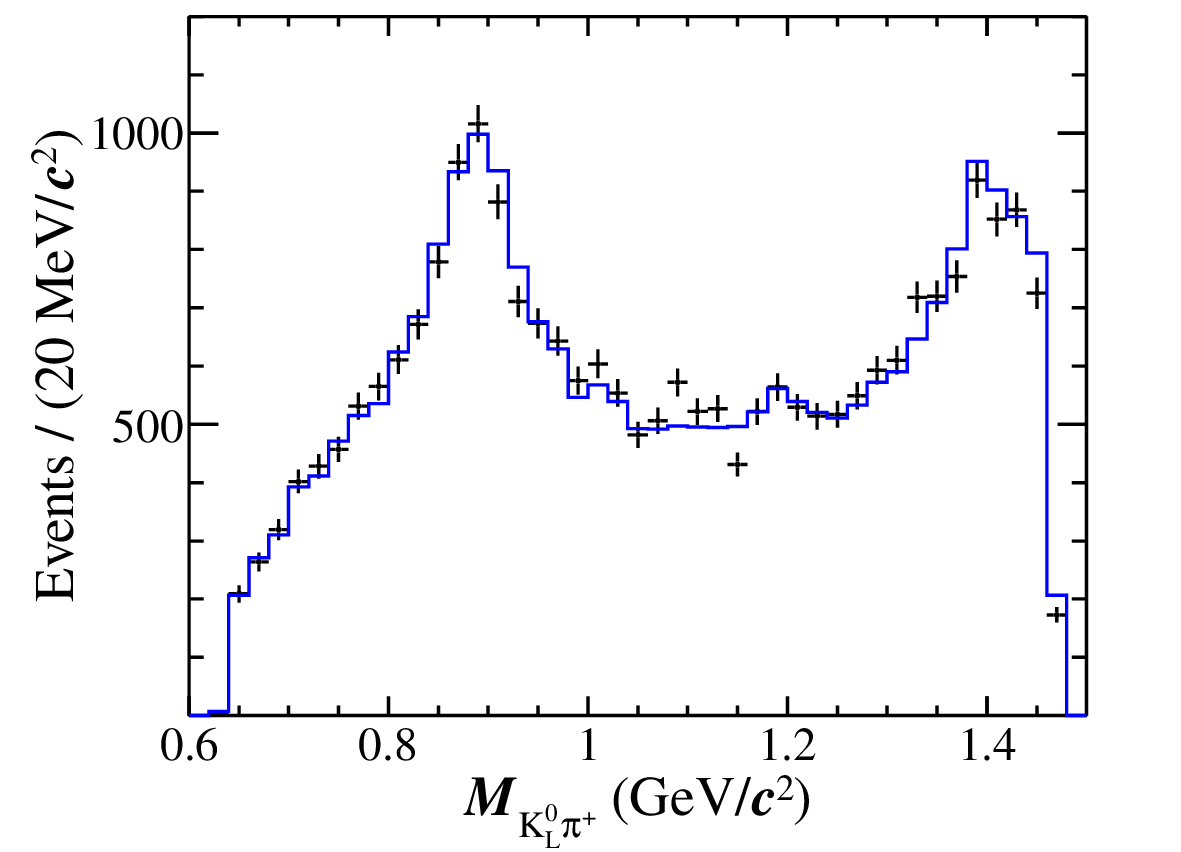}
  \includegraphics[width=3.0in]{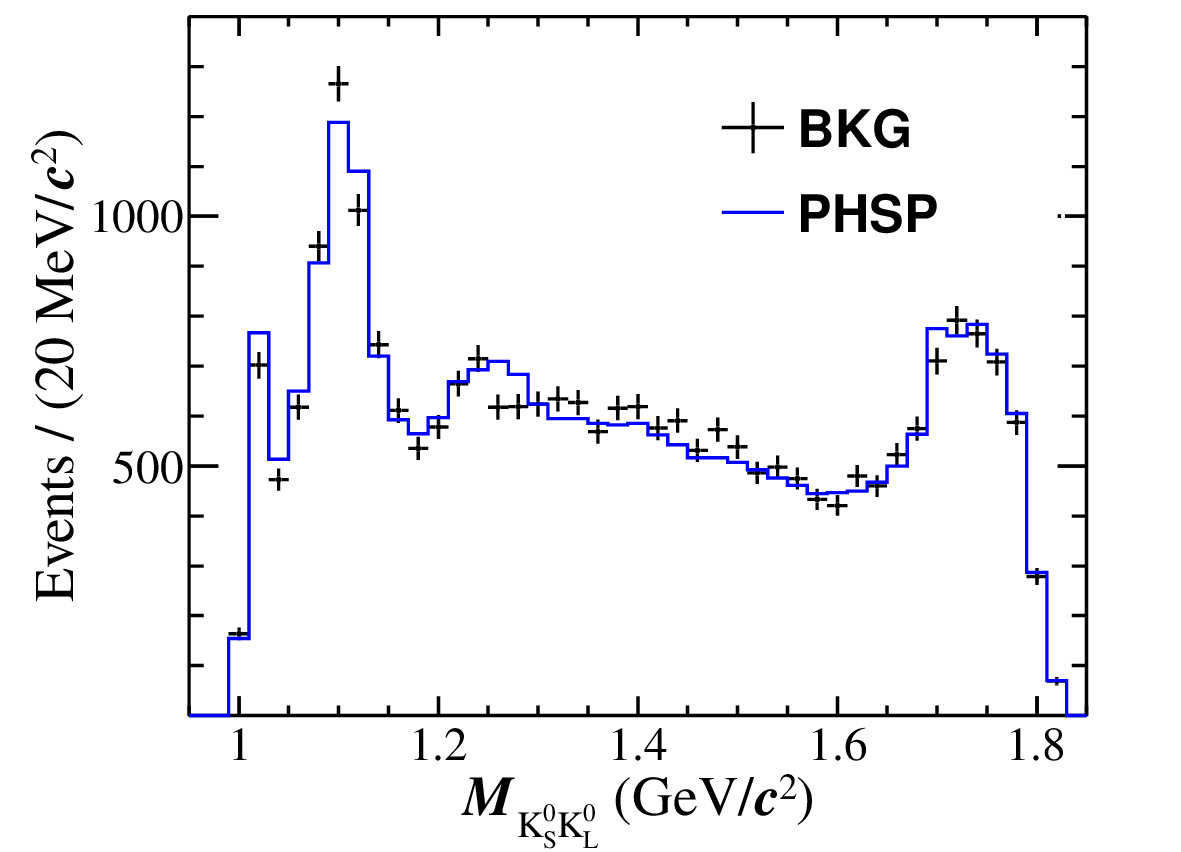}
  \includegraphics[width=3.0in]{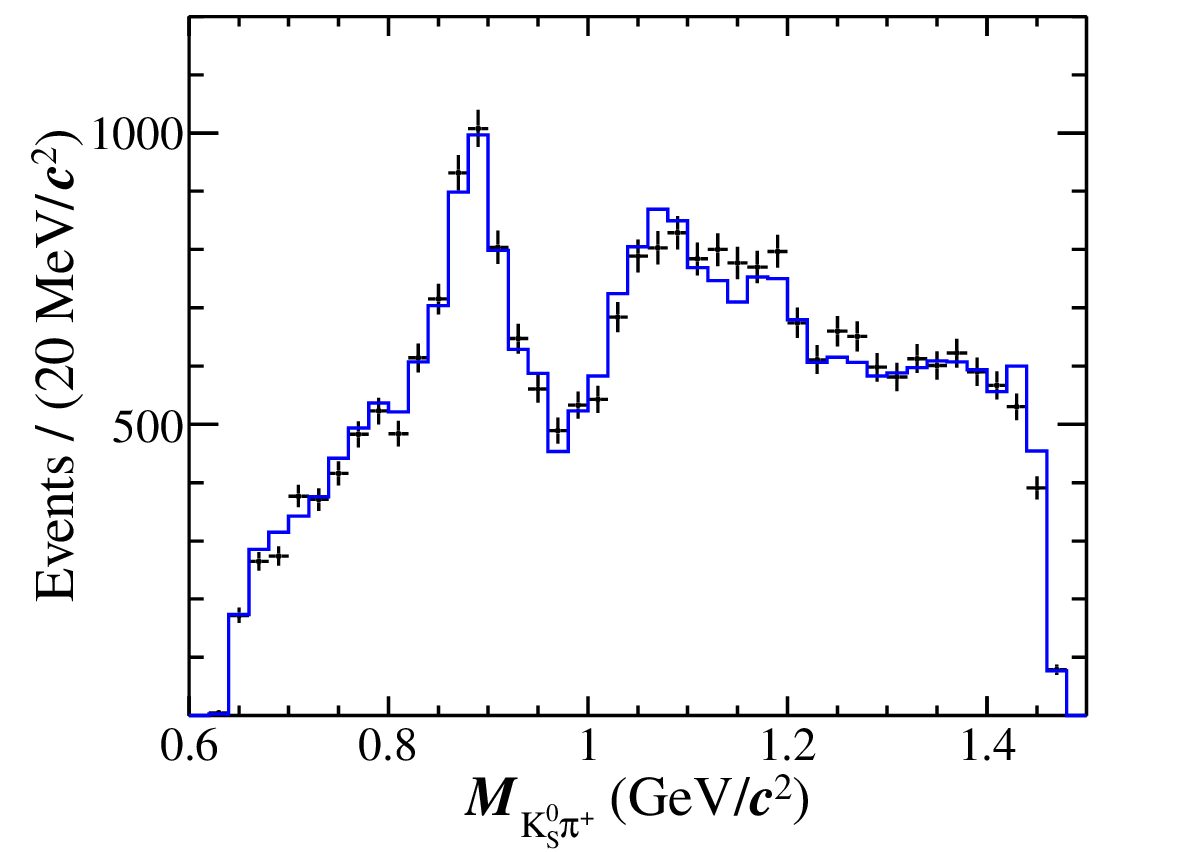}
  \caption{Projections of a background MC sample for
    $D_s^+ \to K_S^0 K_L^0 \pi^+$ and the corresponding background shape obtained from XGBoost trained on that sample. Background shape projections are made by drawing the distributions of a reconstructed MC sample with $P_{\rm {BKG}}/P_{\rm {PHSP}}$ as weight.}
  \label{fig:XGboost}
\end{figure}

To capture the complex multidimensional correlations and dynamic structures within the phase space, a comprehensive set of kinematic variables is provided to the XGBoost algorithm. To ensure optimal performance of the decision trees, the number of input features typically exceeds the absolute independent kinematic degrees of freedom of the decay.
As an illustrative example, Fig.~\ref{fig:XGboost} compares the kinematic projections of a simulated background sample for $D_s^+ \to K_S^0 K_L^0 \pi^+$~\cite{lihui_Dsksklpi} against the learned background distribution $\mathcal{B}_\epsilon$ modeled by the XGBoost classifier.

\subsection{Projection}
Visualizing the agreement between the data and the fitted model through one-dimensional projections onto physical observables---such as invariant masses, decay angles, and particle momenta---provides the most intuitive means to assess fit quality and elucidate the underlying physics. However, analytically constructing these expected projections directly from the fitted parameters is highly nontrivial. The dynamical amplitude must be convolved with the kinematic \textcolor{red}{PHSP} factor $R(p)$ and the detector efficiency $\epsilon(p)$. Unfortunately, explicit analytical forms for these detector and kinematic effects are generally intractable and highly dependent on the choice of coordinates.

In practice, this challenge is circumvented by utilizing a fully reconstructed PHSP MC sample. The expected distribution of the fit result for any given physical observable is constructed by filling a histogram with the reconstructed PHSP MC events, where each event is weighted by its fitted dynamical amplitude squared, $|\mathcal{M}(p)|^2$. Because the reconstructed PHSP MC sample inherently encapsulates both the kinematic PHSP boundaries and the detector efficiency, these critical effects are naturally integrated into the resulting projected distributions without requiring explicit analytical modeling.

\subsection{Fit fraction}
The raw outputs of an amplitude analysis are the relative magnitudes and phases of the intermediate dynamic amplitudes. However, these parameters inherently depend on the chosen normalization and phase conventions of the specific amplitude formalism. Consequently, variations in the amplitude parametrization can significantly alter the fitted parameter values. To enable robust comparisons with independent measurements and to provide meaningful inputs for theoretical phenomenologists, one must extract formalism-independent physical quantities. These quantities, referred to as fit fractions, represent the relative contribution of each individual intermediate process to the total multibody decay rate. Because of quantum interference between intermediate decay channels, the sum of all fit fractions does not necessarily equal unity; the sum will be less than unity in the presence of net constructive interference, and greater than unity for net destructive interference.

The specific fit fraction for the $n^{\mathrm{th}}$ intermediate process is defined as
\begin{equation}\label{eq:FF}
    \mathrm{FF}_{n} = \frac{\int|c_n\mathcal{A}_n(p)|^2 R(p)\mathrm{d}p}{\int|\mathcal{M}(p)|^2 R(p)\mathrm{d}p}.
\end{equation}
Because fit fractions represent pure, post-decay physical quantities, this definition intentionally isolates the underlying dynamics from detector acceptance and resolution effects, thereby fundamentally distinguishing it from the experimentally measured signal PDF, $f_S(p)$. In practice, the integral in Eq.~(\ref{eq:FF}) is evaluated numerically via MC integration using a \textit{generator-level} PHSP MC sample (i.e., prior to any detector simulation):
\begin{equation}
    \mathrm{FF}_{n} \approx \frac{\sum_{k=1}^{N_{\mathrm{gen}}} |c_n \mathcal{A}_n(p_k)|^2}{\sum_{k=1}^{N_{\mathrm{gen}}} |\mathcal{M}(p_k)|^2},
\end{equation}
where $N_{\mathrm{gen}}$ is the total number of generator-level PHSP MC events, and $p_k$ denotes the kinematics of the $k^{\mathrm{th}}$ generated event. The interference fraction between the $n^{\mathrm{th}}$ and $m^{\mathrm{th}}$ amplitudes is derived analogously:
\begin{equation}
    \mathrm{IN}_{nm} \approx \frac{\sum_{k=1}^{N_{\mathrm{gen}}} 2\mathrm{Re}\left[c_{n}c^{*}_{m}\mathcal{A}_{n}(p_k)\mathcal{A}^{*}_{m}(p_k)\right]}{\sum_{k=1}^{N_{\mathrm{gen}}} |\mathcal{M}(p_k)|^2}. \label{interferenceFF-Definition}
\end{equation}

Evaluating the statistical uncertainties of these fit fractions is highly complex, as analytically propagating the uncertainties from the fitted magnitudes and phases is practically unfeasible due to severe nonlinearities and parameter correlations. The standard approach to address this is to perform MC sampling of the fit parameters based on their full covariance matrix (obtained from the fit convergence). This pseudo-experiment procedure generates an empirical distribution for each fit fraction. Typically, these distributions are fitted with a Gaussian function, and the resulting width is assigned as the statistical uncertainty. 

However, it is crucial to recognize that these distributions are not guaranteed to be strictly Gaussian. Strong interference effects or proximity to physical boundaries (e.g., fit fractions near 0\% or 100\%) can heavily skew the distributions. In such asymmetric scenarios, a Gaussian approximation is fundamentally inadequate. An asymmetric Gaussian or a Poisson distribution would be more appropriate.

\section{Amplitude formalism}\label{sec:formalism}
The mathematical formulation and coordinate representation of decay amplitudes are inherently dictated by the underlying physical dynamics of the specific process. This section outlines the formalisms implemented for both hadronic and semileptonic charmed-meson decays at BESIII, including the established parametrizations of intermediate resonance propagators.

\subsection{Hadronic decays}
Amplitude analyses of hadronic charmed-meson decays at BESIII primarily employ the isobar model within the covariant tensor formalism~\cite{Zou:2002ar}, utilizing the four-momenta of the final-state particles as the fundamental kinematic variables. In the isobar model, a multibody decay is conceptualized as a coherent sum of various intermediate quasi-two-body transitions (see Eq.~(\ref{eq:total_amplitude})). 

In a three-body decay, the topological structure typically proceeds via the initial $D$ meson decaying into an intermediate resonance and a bachelor particle, with the resonance subsequently decaying into the remaining two final-state particles. The dynamic amplitude $\mathcal{A}_n$ for such an intermediate process is modeled as
\begin{equation}
  \mathcal{A}_n = P_n S_n F_n^r F_n^{D},
\end{equation}
where $S_n$ represents the spin-projection factor, $F_n^r$ and $F_n^{D}$ are the Blatt-Weisskopf barrier factors for the intermediate resonance and the mother $D$ meson, respectively, and $P_n$ is the resonance propagator that mathematically describes its mass lineshape. 

For four-body decays, the intermediate processes are generally classified into two topological categories: quasi-two-body and cascade. In a quasi-two-body process, the $D$ meson decays into two primary resonance states, each of which subsequently decays into two final-state particles. In a cascade process, the $D$ meson decays into a primary resonance and a bachelor particle; this primary resonance subsequently decays into a secondary resonance and another final-state particle, and the secondary resonance ultimately decays into the final particle pair. In both topologies, the amplitude $\mathcal{A}_n$ is parametrized as
\begin{equation}
  \mathcal{A}_n = P_n^{r_1} P_n^{r_2} S_n F_n^{r_1} F_n^{r_2} F_n^{D},
\end{equation}
where the superscripts $r_1$ and $r_2$ denote the first and second intermediate resonances, respectively. Explicit formulations of the spin factors and Blatt-Weisskopf barriers for arbitrary spin configurations are detailed in Ref.~\cite{Zou:2002ar}. The distinct propagator forms for frequently observed resonances are outlined in Sec.~\ref{sec:propagator}. To satisfy Bose symmetry, the total amplitude $\mathcal{A}_n$ must be explicitly symmetrized under the exchange of any identical final-state bosons. Furthermore, assuming strict $CP$ conservation, the amplitude for a $\bar{D}$ decay is mathematically identical to the $D$ decay amplitude evaluated at the $CP$-conjugate phase-space point. In practical data analysis, this implies that when fitting a $\bar{D}$ data sample, the spatial momenta ($\vec{p}$) of all final-state particles must be inverted ($\vec{p} \to -\vec{p}$) prior to amplitude evaluation.

Any combination of two or three final-state particles can theoretically form a resonant state, which may manifest as a scalar, pseudoscalar, vector, axial-vector, or tensor. However, the physical realization of these intermediate processes is strictly constrained by fundamental quantum selection rules, predominantly angular-momentum conservation. While the initial weak $D$-meson decay intrinsically violates parity, the subsequent resonance decays proceed via strong or electromagnetic interactions where parity is strictly conserved. By rigorously examining the quantum numbers ($J^{PC}$) of the intermediate states, one can systematically deduce the allowed and forbidden transition paths. In practice, a comprehensive suite of kinematically allowed intermediate processes must be empirically evaluated in the fit, utilizing the specific spin factors and Blatt-Weisskopf barriers appropriate for the corresponding orbital angular momenta and resonance species.

\subsection{Semileptonic decays}
The theoretical formulation of semileptonic decays naturally factorizes into a leptonic current and a hadronic current. The dynamics of these two currents can be rigorously separated because there are no final-state strong interactions between the leptonic and hadronic systems~\cite{Zhang:2023nnn}. Consequently, the differential decay amplitude for a $D \to M_{1}M_{2}\ell^+\nu_{\ell}$ transition (where $M_{1,2}$ denote mesons and $\ell=e,\mu$) is naturally parametrized by five independent kinematic variables: the squared invariant masses of the hadronic ($m^2$) and leptonic ($q^2$) systems, their respective helicity angles ($\theta_M$ and $\theta_\ell$), and the angle ($\chi$) between their respective decay planes. Squaring the amplitude and incorporating the phase-space kinematics yields the fully differential decay rate (analogous to the $|\mathcal{M}|^2 R(p)\mathrm{d}p$ term in Eq.~(\ref{eq:likelihood_SB})), which is expressed as
\begin{equation}
  \mathrm{d}\Gamma = \frac{G_F^2|V_{cq}|^2}{(4\pi)^6 m_{D}^3}X\beta_{M}\beta_{\ell}\mathcal{I}(m^2, q^2, \theta_M, \theta_\ell, \chi)\mathrm{d}m^2 \mathrm{d}q^2 \mathrm{d}\!\cos\theta_{M}\mathrm{d}\!\cos\theta_\ell \mathrm{d}\chi\,.
\end{equation}
In this expression, $X\beta_{M}\beta_{\ell}$ represents the kinematic phase-space factor. Here, $X=p_{MM}m_{D}$, where $p_{MM}$ is the magnitude of the three-momentum of the $M_1M_2$ system evaluated in the $D$-meson rest frame, and $m_{D}$ is the $D$-meson mass. The factors $\beta_{M}=2p_{M}/m$ and $\beta_{\ell}=2p_{\ell}/q$ incorporate the momentum magnitudes $p_{M}$ and $p_{\ell}$ of $M_{1}$ and $\ell^+$ evaluated in their respective $M_{1}M_{2}$ and $\ell^{+}\nu_{\ell}$ center-of-mass frames. 

The decay intensity $\mathcal{I}$ (corresponding to the $|\mathcal{M}|^2$ term in Eq.~(\ref{eq:likelihood_SB})) contains the core dynamic information. It is conventionally decomposed in terms of the angular variables $\cos\theta_{\ell}$ and $\chi$ to mathematically isolate the substructure of the hadronic system. This intensity encapsulates the complex hadronic form factors, which are systematically expanded in partial waves according to the angular momentum of the $M_1M_2$ pair. Under the assumption of $CP$ conservation, the amplitude for the charge-conjugate $\bar{D}$ decay is obtained by reversing the sign of the azimuthal angle ($\chi \to -\chi$), while the other four kinematic variables remain invariant. A comprehensive parametrization of this theoretical framework is detailed in Ref.~\cite{Zhang:2023nnn}. The specific parametrizations of the intermediate hadronic resonance propagators are discussed in the subsequent section.

\subsection{Propagator}\label{sec:propagator}
Propagators parameterize mass lineshapes of intermediate resonances. Choice of propagator model for a given resonance depends on its width, its proximity to decay thresholds, and possible presence of overlapping states with the same quantum numbers. This section summarizes the parameterizations employed in amplitude analyses of charmed meson decays.

\subsubsection{Relativistic Breit-Wigner}
For most isolated resonances that are narrow and far from decay thresholds, a relativistic Breit-Wigner (RBW) propagator provides an adequate description~\cite{Jackson:1964zd}. Resonances commonly parameterized in this way include $\omega$, $\phi$, $b_1(1235)$, $a_1(1260)$, $f_2(1270)$,  $a_2(1320)$, $f_0(1370)$, $\eta(1405)$, $a_0(1450)$, $f_0(1500)$, $K^*(892)$, $K_1(1270)$, $K_1(1400)$, $K_2^*(1430)$, etc.

The general form of a RBW propagator is
\begin{equation}
P(s) = \frac{1}{m_0^2 - s - i m_0 \Gamma(s)}\,,
\label{eq:rbw}
\end{equation}
where $s = m^2$ is the invariant mass squared of the decay products. For a two-body decay, the energy-dependent width is given by
\begin{equation}
\Gamma(s) = \Gamma_0 \frac{m_0}{\sqrt{s}} \left( \frac{q}{q_0} \right)^{2L+1} \frac{F_L(q)^2}{F_L(q_0)^2}\,.
\label{eq:rbw_width}
\end{equation}
Here, $m_0$ and $\Gamma_0$ are the mass and width of the intermediate resonance, which can be fixed to their known values~\cite{PDG}. The quantity $q$ is the magnitude of the breakup momentum of the daughter particles in the resonance rest frame, $q_0 = q(s=m_0^2)$, and $F_L$ is the Blatt-Weisskopf barrier factor for orbital angular momentum $L$; their explicit definitions can be found in Ref.~\cite{PDG}. For axial-vector mesons such as $a_1(1260)$, $K_1(1270)$, and $K_1(1400)$, which decay predominantly through three-body processes, a more general mass-dependent width $\Gamma(s)$ should be used; further details can be found in Ref.~\cite{Argent}.

\subsubsection{Gounaris-Sakurai}
For broad vector resonances, such as $\rho(770)$ and $\rho(1450)$, a simple RBW form fails to describe the lineshape accurately near threshold. In these cases, the Gounaris-Sakurai (GS) parametrization~\cite{GS} is adopted, which imposes analyticity constraints on the $\pi\pi$ P-wave amplitude:
\begin{equation}
P_{\text{GS}}(s) = \frac{1 + d\,\Gamma_0/m_0}{m_0^2 - s + f(s) - i m_0 \Gamma(s)}.
\label{eq:gs}
\end{equation}
The function $f(s)$ and the constant $d$ are defined in Ref.~\cite{GS}; $d = f(0)/(\Gamma_0 m_0)$ is fixed by the normalization at $s=0$. In certain cases, the $\pi^+\pi^-$ mass spectrum in the $\rho(770)$ region cannot be adequately described by the GS lineshape alone, owing to distortions induced by $\rho-\omega$ mass mixing. When these effects are significant, a $\rho$--$\omega$ mixing lineshape~\cite{mix2} should be adopted to account for the interference.

\subsubsection{\texorpdfstring{$\boldsymbol{f_0(500)}$}{f0(500)}}
The $\sigma/f_0(500)$ is a very broad scalar resonance with strong coupling to multiple channels. Its propagator is parameterized following Ref.~\cite{f05001} as
\begin{equation}
P_{f_0(500)}(s) = \frac{1}{m_0^2 - s - i m_0 \Gamma_{\text{tot}}(s)}\,,
\end{equation}
with $\Gamma_{\text{tot}}(s) = g_1 \frac{\rho_{\pi\pi}(s)}{\rho_{\pi\pi}(m_0^2)} + g_2 \frac{\rho_{4\pi}(s)}{\rho_{4\pi}(m_0^2)}$. Here $\rho_{\pi\pi}(s)$ and $\rho_{4\pi}(s)$ are the Lorentz-invariant phase-space factors for the two-pion and four-pion channels, and $g_{1,2}$ are the corresponding coupling constants. Their detailed parametrizations and numerical values are taken from Ref.~\cite{f05002}.

\subsubsection{\texorpdfstring{$\boldsymbol{f_0(980)}$}{f0(980)}}
The $f_0(980)$ couples strongly to $\pi\pi$ and $K\bar{K}$, and lies just below the $K\bar{K}$ mass threshold. A Flatt\'e formula~\cite{Flatte_f0} is therefore used to address the threshold effect:
\begin{equation}
P_{f_0(980)}(s) = \frac{1}{m_0^2 - s - i(g_1\rho_{\pi\pi}(s) + g_2\rho_{K\bar{K}}(s))}\,,
\end{equation}
where $\rho_{\pi\pi}(s)$ and $\rho_{K\bar{K}}(s)$ are the Lorentz-invariant PHSP factors, and $g_{1,2}$ are their coupling constants. Their definitions can be found in Ref.~\cite{Flatte_f0}. Below the $K\bar{K}$ threshold, the analytic continuation $\sqrt{1-4m_K^2/s} \to i\sqrt{4m_K^2/s-1}$ is applied. The parameters can be fixed to the values reported in Ref.~\cite{Flatte_f0}.

\subsubsection{\texorpdfstring{$\boldsymbol{a_0(980)}$}{a0(980)}}
The $a_0(980)$ couples strongly to $\pi\eta$ and $K\bar{K}$, and lies close to the $K\bar{K}$ threshold, requiring a coupled-channel treatment~\cite{BCKa03,BCKa0,Zhang:2022xpf,Zhang:2024myn}. Two parameterizations are considered. The first is a Flatt\'e form~\cite{Flatte_a0}:
\begin{equation}
P_{a_0(980)}(s) = \frac{1}{m_0^2 - s - i \sum_j g_j^2 \rho_j(s)}\,,
\quad j = \pi\eta, K\bar{K}, \pi\eta^\prime\,.
\end{equation}
Here $g_j$ and $\rho_j(s)$ denote the coupling constant and PHSP factor for channel $j$, respectively. This retains only the imaginary part of the self-energy, and is adequate when the PHSP varies slowly and no sharp thresholds lie near the resonance peak.

The second is a dispersive approach~\cite{Bugg08,BESIII_a0}, which includes the full complex self-energy $\Pi_j(s) = \text{Re}\,\Pi_j(s) + i\,\text{Im}\,\Pi_j(s)$:
\begin{equation}
P_{a_0(980)}(s) = \frac{1}{m_0^2 - s - \sum_j g_j^2 \Pi_j(s)}\,.
\end{equation}
The imaginary part is given by $\text{Im}\,\Pi_j(s) = \rho_j(s) F_j^2(s)$, where $F_j(s)$ is a form factor~\cite{Bugg08}, and the real part is obtained from the dispersion relation
\begin{equation}
\text{Re}\,\Pi_j(s) = \frac{1}{\pi}\,\mathcal{P}\!\!\int_{s_{\rm thr}}^{\infty} \frac{\text{Im}\,\Pi_j(s')}{s' - s}\,ds'.
\end{equation}
This formulation naturally accounts for the prominent cusp at the $K\bar{K}$ threshold. The parameters can be fixed to those in Ref.~\cite{BESIII_a0}.

\subsubsection{\texorpdfstring{$\boldsymbol{\pi\pi~S}$-wave}{pi pi S-Wave}}
For the $\pi^+\pi^-$ and $\pi^0\pi^0$ $S$-waves, multiple broad and overlapping resonances appear, and a simple sum of \textcolor{red}{Breit-Wigner} propagators would violate unitarity. In such cases, a $K$-matrix parametrization~\cite{km3,KpiS_1} is adopted. The amplitude is expressed as
\begin{equation}
   \textcolor{red}{ A_i = (\bm{I} - i\bm{K\rho})^{-1}_{ij}P_j\,,}
\end{equation}
\textcolor{red}{where $\bm{I}$ is the identity matrix, $\bm{K}$ is the scattering matrix, and $\bm{\rho}$ is the phase-space matrix.} The indices $i,j$ label the coupled channels: $1 = \pi\pi$, $2 = K\bar{K}$, $3 = 4\pi$, $4 = \eta\eta$, $5 = \eta\eta'$. The production vector $P$ is parametrized as
\begin{equation}
    P_j(s) = f_{1j}^{\rm prod}\frac{1 - s_0^{\rm scatt}}{s - s_0^{\rm scatt}} + \sum_{\alpha}\frac{\beta^{\alpha} g_j^{\alpha}}{m_{\alpha}^2 - s}\,.
\end{equation}
All parameters not explicitly defined here (including the $K$-matrix elements, $f_{1j}^{\rm prod}$, $\beta^{\alpha}$, $s_0^{\rm scatt}$, $g_j^{\alpha}$ and $m_{\alpha}$) are taken from the literature~\cite{km3,KpiS_1}. While the scattering $K$-matrix is usually fixed based on independent scattering data, the production parameters $f_{1j}^{\rm prod}$ and $\beta^{\alpha}$ are process-dependent and left free in the fit.

\subsubsection{\texorpdfstring{$\boldsymbol{K\pi~S}$-wave}{K pi S-Wave}}
For the $K\pi$ $S$-wave, two complementary parametrizations are employed. The LASS model~\cite{KpiS_1} describes the amplitude as a coherent sum of a $K_0^*(1430)$ Breit-Wigner resonance~\cite{PDG} and an effective-range non-resonant component:
\begin{equation}
A(m) = F \sin\delta_F e^{i\delta_F} + R \sin\delta_R e^{i\delta_R} e^{i2\delta_F}\,.
\end{equation}
Here $F$ ($\phi_F$) and $R$ ($\phi_R$) are the magnitudes (phases) for the non-resonant and resonant terms. Their relative phase is fixed by Watson's theorem, making this model well suited for the low-mass region where inelastic channels are negligible. For analyses covering a wider energy range where coupled-channel effects become important, a $K$-matrix model~\cite{KpiS_2} is also employed. This model splits the amplitude into isospin components $\mathcal{A}_{1/2}$ and $\mathcal{A}_{3/2}$, treating resonant and nonresonant contributions on the same footing and guaranteeing unitarity with all relevant coupled channels. The parameters can be cited from the Ref.~\cite{KpiS_3}.

\section{Summary and Discussion}
Amplitude analysis serves as a robust analytical framework that bridges the gap between experimental measurements and theoretical phenomenologies, enabling the extraction of fundamental two-body intermediate dynamics from complex multibody final states. In this work, we have provided a comprehensive review of the amplitude-analysis methodologies employed for charmed-meson decays at the BESIII experiment, with a strong emphasis on practical experimental implementation. We detailed the construction of the likelihood functions, the utilization of MC integration for strict normalization, the treatment of detector efficiencies and finite resolutions, and the modeling of backgrounds via multidimensional reweighting techniques. Furthermore, we outlined the specific amplitude formalisms governing both hadronic and semileptonic decays, including the standard parametrizations for intermediate resonance propagators.

The practical execution of amplitude analysis relies critically on dedicated MC simulations. The selection of a specific MC sample is intrinsically tied to the analytical task, guided by a clear functional mapping: a generator-level PHSP MC sample strictly represents the pure kinematic phase-space boundary; a fully reconstructed PHSP MC sample naturally folds in the detector acceptance and efficiency; and a reconstructed signal MC sample further encapsulates the empirical detector resolution effects. Consequently, mapping a theoretical amplitude model onto observable data projections necessitates a reconstructed MC sample, whereas the extraction of purely physical fit fractions for intermediate processes strictly requires a generator-level PHSP MC sample.

Leveraging these comprehensive methodologies, the BESIII Collaboration has determined the branching fractions for key charmed-meson decays~\cite{lihui_Dsksklpi,BESIII:2024muy,yangliping_kspi0pi0,BESIII:2018mwk,BESIII:2024ncc,BESIII:2023mie,BESIII:2022vaf,BESIII:2023qgj}, including pivotal channels such as $D \to K^{*}\pi$~\cite{yangliping_kspi0pi0, BESIII:2024ncc} and $D_s \to \phi\pi$~\cite{lihui_Dsksklpi,BESIII:2024muy}. These results provide crucial experimental constraints on the nonperturbative dynamics of Quantum Chromodynamics~(QCD). \textcolor{red}{The branching fractions of decays involving scalar and axial-vector mesons have also been precisely measured~\cite{lihui_Dskskspi,lihui_Dskskpi0,BESIII:2019jjr, BESIII:2026mbo, BESIII:2026mtz, BESIII:2025wmd}. Theoretical predictions for decays involving scalar mesons are highly sensitive to their assumed internal quark structures, such as conventional $q\bar{q}$ states versus tetraquark configurations~\cite{BCKa02,haiyang_4k,BCKa03,BCKa0}. For decays involving axial-vector $K_1$ mesons, predictions vary widely as well, due to strong dependence on both the chosen theoretical approach and the poorly constrained $K_1$ mixing angle~\cite{Shi:2023kiy}. These theoretical difficulties and the scarcity of reliable predictions make experimental inputs essential for clarifying the underlying dynamics. Beyond branching fractions, amplitude analyses have enabled the extraction of complex polarization observables in $D \to VV$ decays~\cite{zengx_DstoKpipipi0,zengx_DtoKpipipi0,BESIII:2025nou} and the precise determination of the $CP$-even fractions in $D^0$ multibody decays.} Furthermore, systematic comparisons across these multibody channels enable independent determinations of absolute $\phi$-meson decay branching fractions~\cite{lihui_Dsksklpi,BESIII:2024muy, BESIII:2026lnl}. By performing simultaneous amplitude fits across multiple coupled decay channels, we uniquely probe $K_S^0$--$K_L^0$ asymmetries~\cite{lihui_Dsksklpi}, evaluate $U$-spin symmetry breaking, and explore fundamental quantum correlations within the neutral $D^0$ system.

The BESIII experiment has accumulated unprecedented \textcolor{red}{charmonium} threshold data samples, corresponding to an integrated luminosity of \textcolor{red}{$20.3~\mathrm{fb}^{-1}$} at $\sqrt{s}=3.773~\mathrm{GeV}$~\cite{BESIII:2024lbn} and an additional $7.33~\mathrm{fb}^{-1}$ in the energy range between $4.128$ and $4.226~\mathrm{GeV}$. Capitalizing on these massive datasets, a new generation of high-precision amplitude analyses is currently underway. We anticipate a wealth of groundbreaking results, including unparalleled precision in branching fractions, deeper resolution of broad resonant structures and polarizations, and stringent tests of fundamental symmetries, which will collectively and significantly advance our global understanding of charm-decay dynamics.

\section*{ACKNOWLEDGMENTS}
H. Z. and B.-C. K. were supported in part by National Natural Science Foundation of China (NSFC) under Contracts No. 12192263, Joint Large-Scale Scientific Facility Fund of the NSFC and the Chinese Academy of Sciences under Contract No. U2032104, and the Excellent Youth Foundation of Henan Scientific Commitee under Contract No. 242300421044; C.~Y. G. and L.~Y. D. were supported in part by NSFC under Contracts No. 12192262; Y. L. was supported in part by NSFC under Contracts No.  12575095; H. L. and M.~G. Z were supported in part by NSFC under Contracts No.  123B2077, 12035009.


\begin{thebibliography}{99}






\bibitem{PDG}
S. Navas \textit{et al.} (Particle Data Group), {\href{https://doi.org/10.1103/PhysRevD.110.030001}{Phys. Rev. D \textbf{110}, 030001 (2024).}}

\bibitem{Achasov:2017edm}
N.~N.~Achasov and G.~N.~Shestakov,
\href{https://journals.aps.org/prd/abstract/10.1103/PhysRevD.96.036013}{Phys. Rev. D \textbf{96}, 036013 (2017).}

  \bibitem{BESIII:2024lbn}
M. Ablikim {\it et al.} (BESIII Collaboration), \href{https://arxiv.org/abs/2406.05827}{Chin. Phys. C {\bf 48}, 123001 (2024).}

\bibitem{geant4} S.~Agostinelli {\it et al.} (GEANT4 Collaboration),\href{https://inspirehep.net/files/6c9c0b62bbc8dc0401fca11a5fe5c87c} {Nucl. Instrum. Meth. A {\bf 506}, 250 (2003).}

\bibitem{Liu:2019huh} B.~Liu, X.~Xiong, G.~Hou, S.~Song and L.~Shen, \href{https://www.epj-conferences.org/articles/epjconf/abs/2019/19/epjconf_chep2018_06033/epjconf_chep2018_06033.html}{EPJ Web Conf. \textbf{214}, 06033 (2019).}
  
\bibitem{XGboost} X. D. Team, Xgboost official website, Available at:  \href{https://xgboost.readthedocs.io/en/latest/}{https://xgboost.readthedocs.io/en/latest/}

\bibitem{lihui_Dsksklpi} M.~Ablikim {\it et al.} (BESIII Collaboration), \href{https://doi.org/10.1103/6py9-h8qv} {Phys. Rev. Lett. {\bf 135}, 161902 (2025).}

\bibitem{Zou:2002ar}
B. S. Zou, and D. V. Bugg, {\href{https://link.springer.com/article/10.1140/epja/i2002-10135-4}{Eur. Phys. J. A \textbf{16}, 537-547 (2003)}}
  



\bibitem{Zhang:2023nnn}
H.~Zhang, B.~C.~Ke, Y.~Yu and E.~Wang, {\href{https://iopscience.iop.org/article/10.1088/1674-1137/acc642}{Chin. Phys. C \textbf{47}, 063101 (2023)}}

\bibitem{Jackson:1964zd}
J.~D.~Jackson, \href{https://link.springer.com/article/10.1007/BF02750563}{Nuovo Cim. \textbf{34}, 1644-1666 (1964).}

\bibitem{Argent}
P. d'Argent \textit{et al.} (CLEOc Collaboration), {\href{https://doi.org/10.1007/JHEP05(2017)143}{JHEP \textbf{05}, 143 (2017)}}

\bibitem{GS}
G. J. Gounaris and J. J. Sakurai, {\href{https://doi.org/10.1103/PhysRevLett.21.244}{Phys. Rev. Lett. \textbf{21}, 244 (1968)}}


\bibitem{mix2}
R. R. Akhmetshin \textit{et al.} (CMD-2 Collaboration), {\href{https://doi.org/10.1016/S0370-2693(02)01168-1}{Phys. Lett. B \textbf{527}, 161 (2002)}}

\bibitem{f05001}
D. V. Bugg, A. V. Sarantsev, and B. S. Zou, {\href{https://www.sciencedirect.com/science/article/pii/0550321396001666?via%3Dihub}{Nucl. Phys. B \textbf{471}, 59 (1996)}}     

\bibitem{f05002}
M. Ablikim \textit{et al.} (BESIII Collaboration), {\href{https://www.sciencedirect.com/science/article/pii/S0370269304011384?via%3Dihub}{Phys. Lett. B \textbf{598}, 149 (2004)}}                                             

\bibitem{Flatte_f0}
M. Ablikim \textit{et al.} (BES Collaboration), {\href{https://www.sciencedirect.com/science/article/pii/S0370269304017265?via%3Dihub}{Phys. Lett. B \textbf{607}, 243 (2005).}}                                                


\bibitem{BCKa03} Y.~K.~Hsiao, S. Q. Yang, W. J. Wei, and B. C. Ke,
\href{https://link.springer.com/article/10.1007/JHEP12(2024)226} {JHEP \textbf{12}, 226 (2025).}

\bibitem{BCKa0} Y.~K.~Hsiao, Y.~Yu, and B.~C.~Ke,
  \href{https://doi.org/10.1140/epjc/s10052-020-08468-9} {Eur. Phys. J. C \textbf{80}, 895 (2020).}
  
\bibitem{Zhang:2022xpf} H. Zhang, Y. H. Lyu, L. J. Liu, and E. Wang,
  \href{https://iopscience.iop.org/article/10.1088/1674-1137/acb3b3}{Chin. Phys. C {\bf 47} 043101 (2023).}
  
\bibitem{Zhang:2024myn}
X.H.~Zhang, H.~Zhang, B.~C.~Ke, {\it et al.} 
\href{https://journals.aps.org/prd/abstract/10.1103/PhysRevD.110.114050} {Phys. Rev. D \textbf{110}, 114050 (2024).}

\bibitem{Flatte_a0}
G. S. Adams \textit{et al.} (CLEO Collaboration), {\href{https://doi.org/10.1103/PhysRevD.84.112009}{Phys. Rev. D \textbf{84}, 112009 (2011)}}

\bibitem{Bugg08}
D. V. Bugg, {\href{https://doi.org/10.1103/PhysRevD.78.074023}{Phys. Rev. D \textbf{78}, 074023 (2008)}}    

\bibitem{BESIII_a0}
M. Ablikim \textit{et al.} (BESIII Collaboration), {\href{https://doi.org/10.1103/PhysRevD.95.032002}{Phys. Rev. D \textbf{95}, 032002 (2017)}}

\bibitem{km3}
V. V. Anisovich and A. V. Sarantsev, {\href{https://link.springer.com/article/10.1140/epja/i2002-10068-x}{Eur. Phys. J. A \textbf{16}, 229 (2003)}}

\bibitem{KpiS_1}
I. Adachi \textit{et al.} (BABAR and Belle Collaborations), {\href{https://doi.org/10.1103/PhysRevD.98.112012}{Phys. Rev. D \textbf{98}, 112012 (2018)}}

\bibitem{KpiS_2}
R. Aaij \textit{et al.} (LHCb Collaboration), {\href{https://doi.org/10.1140/epjc/s10052-018-5758-4}{Eur. Phys. J. C \textbf{78}, 443 (2018)}}

\bibitem{KpiS_3}
J. M. Link \textit{et al.} (FOCUS Collaboration), {\href{https://www.sciencedirect.com/science/article/pii/S0370269307007927?via%3Dihub}{Phys. Lett. B \textbf{653}, 1 (2007)}}  
\bibitem{BESIII:2018mwk}
M. Ablikim {\it et al.} (BESIII Collaboration), \href{https://doi.org/10.1103/PhysRevD.99.091101} {Phys. Rev. D {\bf 99}, 091101 (2019).}

\bibitem{BESIII:2024ncc}
M. Ablikim {\it et al.} (BESIII Collaboration), \href{https://journals.aps.org/prd/abstract/10.1103/PhysRevD.110.092006} {Phys. Rev. D {\bf 110}, 092006 (2024).}

\bibitem{BESIII:2023mie}
M. Ablikim {\it et al.} (BESIII Collaboration), \href{https://journals.aps.org/prd/abstract/10.1103/PhysRevD.107.052010} {Phys. Rev. D {\bf 107}, 052010 (2023).}

\bibitem{BESIII:2022vaf}
M. Ablikim {\it et al.} (BESIII Collaboration), \href{https://doi.org/10.1007/JHEP03(2026)060}
{JHEP \textbf{08} (2022) 196}.

\bibitem{BESIII:2024muy} M.~Ablikim {\it et al.} (BESIII Collaboration), \href{https://journals.aps.org/prl/abstract/10.1103/PhysRevLett.134.011904} {Phys. Rev. Lett. {\bf 134}, 011904 (2025).}

  
\bibitem{BESIII:2023qgj}
M. Ablikim {\it et al.} (BESIII Collaboration), \href{https://link.springer.com/article/10.1007/JHEP09(2023)077}
{JHEP \textbf{09}, 077 (2023)}.

\bibitem{yangliping_kspi0pi0}
M. Ablikim {\it et al.} (BESIII Collaboration), \href{https://doi.org/10.1007/JHEP03(2026)060}
{JHEP \textbf{03} (2026) 060}.

\bibitem{lihui_Dskskspi}
M. Ablikim {\it et al.} (BESIII Collaboration), \href{https://journals.aps.org/prd/abstract/10.1103/PhysRevD.105.L051103} {Phys. Rev. D {\bf 105}, L051103 (2022).}

\bibitem{lihui_Dskskpi0}
 M.Ablikim {\it et al.} (BESIII Collaboration), \href{https://journals.aps.org/prl/abstract/10.1103/PhysRevLett.129.182001} {Phys. Rev. Lett. {\bf 129}, 182001 (2022).}

\bibitem{BESIII:2019jjr}
M. Ablikim {\it et al.} (BESIII Collaboration), \href{https://doi.org/10.1103/PhysRevLett.123.112001} {Phys. Rev. Lett. {\bf 123}, 1112001 (2019).}

\bibitem{BESIII:2026mbo}
M.~Ablikim \textit{et al.} (BESIII Collaboration), \href{https://arxiv.org/abs/2604.10444}{[arXiv:2604.10444 [hep-ex]].}

\bibitem{BESIII:2026mtz}
M.~Ablikim \textit{et al.} (BESIII Collaboration), \href{https://arxiv.org/abs/2603.18521}{[arXiv:2603.18521 [hep-ex]].}

\bibitem{BESIII:2025wmd}
M.~Ablikim \textit{et al.} (BESIII Collaboration), \href{https://arxiv.org/abs/2512.23389}{[arXiv:2512.23389 [hep-ex]].}

\bibitem{BCKa02} Y.~Yu, Y.~K.~Hsiao, and B.~C.~Ke,
  \href{https://doi.org/10.1140/epjc/s10052-021-09895-y} {Eur. Phys. J. C \textbf{81}, 1093 (2021).}
\bibitem{haiyang_4k}
H.~Y.~Cheng and C.~W.~Chiang,
\href{https://journals.aps.org/prd/abstract/10.1103/PhysRevD.110.094052} {Phys. Rev. D \textbf{110}, 094052 (2024). }



\bibitem{Shi:2023kiy} Y.~J.~Shi, J.~Zeng and Z.~F.~Deng, \href{https://journals.aps.org/prd/abstract/10.1103/PhysRevD.109.016027}{Phys. Rev. D \textbf{109}, no.1, 016027 (2024).}

\bibitem{zengx_DstoKpipipi0}
M. Ablikim {\it et al.} (BESIII Collaboration), \href{https://link.springer.com/article/10.1007/JHEP09(2022)242}
{ JHEP 09 (2022) 242}.

\bibitem{zengx_DtoKpipipi0}
M. Ablikim {\it et al.} (BESIII Collaboration), \href{https://link.springer.com/article/10.1007/JHEP05(2025)195}
{ JHEP 05 (2025) 195}.

\bibitem{BESIII:2025nou} M.~Ablikim {\it et al.} (BESIII Collaboration), \href{https://journals.aps.org/prl/abstract/10.1103/PhysRevLett.134.201902} {Phys. Rev. Lett. {\bf 134}, 201902 (2025).}

\bibitem{BESIII:2026lnl} M.~Ablikim {\it et al.} (BESIII Collaboration), \href{https://arxiv.org/abs/2605.11464}{arXiv:2605.11464 [hep-ex].}




\end{thebibliography}
\end{document}